\documentstyle[12pt]{article}
\def\laq{\raise 0.4 ex \hbox{$<$}\kern -0.8 em\lower 0.62 ex\hbox{$\sim$}}
\def\gaq{\raise 0.4 ex \hbox{$>$}\kern -0.7 em\lower 0.62 ex\hbox{$\sim$}}
 
\def\beq{\begin{equation}}
\def\eeq{\end{equation}}
\def\bea{\begin{eqnarray}}
\def\eea{\end{eqnarray}}
\def\bq{\begin{quote}}
\def\eq{\end{quote}}

\def\frac#1#2{{\textstyle{{#1}\over {#2}}}}

\def\lsim{\mathrel{\rlap{\lower4pt\hbox{\hskip1pt$\sim$}}
    \raise1pt\hbox{$<$}}}
\def\gsim{\mathrel{\rlap{\lower4pt\hbox{\hskip1pt$\sim$}}
    \raise1pt\hbox{$>$}}}
\def\sqr#1#2{{\vcenter{\vbox{\hrule height.#2pt
         \hbox{\vrule width.#2pt height#1pt \kern#1pt
         \vrule width.#2pt}
         \hrule height.#2pt}}}}

\def\CQG{{\it Class. Quantum Gravity} }

\def\GRG{{\it Gen. Relativity and Gravitation} }

\def\NP{{\it Nucl. Phys.} }
\def\PL{{\it Phys. Lett.} }
\def\PR{{\it Phys. Rev.} }
\def\PRL{{\it Phys. Rev. Lett.} }
\def\PRTS{{\it Phys. Rep.} }

\def\gappeq{\mathrel{\rlap {\raise.5ex\hbox{$>$}}
{\lower.5ex\hbox{$\sim$}}}}
 
\def\lappeq{\mathrel{\rlap{\raise.5ex\hbox{$<$}}
{\lower.5ex\hbox{$\sim$}}}}

\begin{document}

\begin{flushright}
{DF/IST - 8.2002} \\
{July 2002} \\
\end{flushright}
\vglue 0.5cm
	
\begin{center}
{{\bf  Hypothetical Gravity Control and Implications for Spacecraft Propulsion}\footnote{Based on talks presented at the 
Advanced Propulsion sections of the International Space University Workshop and of the 6th International Symposium ``Propulsion for 
Space Transportation for the XXIst Century'', Versailles, France, May 2002.}\\}

\vspace*{2.0cm}

O. Bertolami$^{(a),}$\footnote{E-mail address: {\tt orfeu@cosmos.ist.utl.pt}} 
and M. Tajmar$^{(b),}$\footnote{E-mail address: {\tt martin.tajmar@arcs.ac.at}}\\

\vglue 0.5 cm

{(a) Instituto Superior T\'ecnico, Departamento de F\'\i sica, 1049-001 Lisboa, Portugal}\\

\vglue 0.3cm

{(b) ARC Seibersdorf research GmbH, Space Propulsion, A-2444 Seibersdorf, Austria}\\

\end{center}

\setlength{\baselineskip}{0.7cm}

\vglue 1cm

\centerline{\bf  Abstract}
\vglue 0.5cm
\noindent
{\bf A scientific analysis of the conditions under which gravity could be 
controlled and the implications that an hypothetical manipulation of gravity 
would have for known schemes of space propulsion 
have been the scope of a recent study carried out for the European Space 
Agency. 
The underlying fundamental physical principles of known theories of gravity 
were analysed and shown that even if gravity could be modified it would bring 
somewhat modest gains in terms of launching of spacecraft and no breakthrough 
for space propulsion.}

\vglue 1cm

\pagestyle{plain}

\setcounter{equation}{0}
\setlength{\baselineskip}{0.7cm}

Putative control of gravity is a topic repeatedly discussed in scientific publications, 
experiments and even patents. 
Aiming to critically assess the existing literature and to reach a set of 
recommendations for future activity, the European Space Agency (ESA) has ordered 
the authors a scientific study. This study comprised a scientific report 
of about 140 pages, a database 
containing about 150 key papers on the subject, as well as information on 
38 scientifically active individuals in the research of new space propulsion schemes and its  
underlying fundamental physical principles \cite{Bertolami1}. 

On quite broad terms, we understand by {\it gravity control}  
any scheme to alter the effective strength 
of the gravitational coupling to matter or that can lead, through the 
intervention of other forces, to a change in the local gravity 
force. For achieving this goal, at least one of the following conditions 
must be fulfilled:

\noindent
{\bf 1) } Existence of a new fundamental interaction of Nature 
so to alter the effective strength of the gravitational coupling to matter. This  
implies in violations of the Weak Equivalence Principle.

\noindent
{\bf 2)} Existence of net forces due to the interplay between gravity and
electrostatic forces in shielded experimental configurations, as found 
in the well-known Schiff-Barnhill effect \cite{Schiff}.

\noindent
{\bf 3)} Analogous effect for magnetic fields in quantum 
materials involving the gravitomagnetic field \cite{DeWitt}.

\noindent
{\bf 4)} Physically altering the properties of the vacuum so to change the relative 
strength of the known fundamental interactions of Nature.

In order to critically examine the existing proposals and to consider lines of 
research that could lead to yet unknown phenomena, a survey on the state of the 
art of the following topics has been carried out:

\begin{itemize}

\item General Relativity

\item Einstein Equivalence Principle:  
Weak Equivalence Principle, Local Lorentz Invariance, 
Local Position Invariance

\item New Interactions of Nature 

\item String/M-Theory

\item Gravitoelectromagnetism in Quantum Materials

\item ``Vacuum Engineering''

\item Gravity-Controlled Propulsion

\end{itemize}

The conclusions of our survey can be summarized as follows:

\begin{itemize}

\item Schemes involving heterodox concepts such as {\it negative masses}, 
{\it negative energy densities}, {\it Mach's principle, warp drive}, etc, 
contradict well tested conventional physical theories and cannot so far 
be seriously considered.

\item Gravitational anomalies related to shielding effects (claimed to be 
observed during total eclipses), reduction 
of weight due to rotation of quantum materials and amplification of 
gravitomagnetic fields are almost entirely ruled out. Experimental 
examination of claims that have not yet been ruled out is recommended through 
specifically designed experiments aimed to measure the gravitomagnetic properties 
of rotating quantum materials \cite{Bertolami1}.

\item The Weak Equivalence Principle holds with great accuracy, actually in 
$5$ parts in $10^{13}$ \cite{Adelberger} (see \cite{Nieto} for an extensive review) 
and expected violations, such as the one emerging in certain string theory 
schemes \cite{Damour}, are far too small, $1$ part in $10^{18}$, to 
lead to any workable scheme to control gravity. 
The same can be stated about Lorentz (see eg. \cite{Bertolami2}) and CPT 
symmetries \cite{Kostelecky}.

\item A new interaction of Nature arising from the exchange of a new intermediate massive  
boson is ruled out for ranges in the interval 
$10^{-4}~m \lsim \lambda \lsim 10^{13}~m$ \cite{Fishbach}. 
Submillimeter range new interactions have been recently very much discussed in the context of 
braneworld scenarios with large extra dimensions \cite{Antoniadis}, but arise also
from assuming the new particle account for vacuum energy density in the Universe 
\cite{Beane,Bertolami3}. Interestingly, this range has recently became accessible to 
experimental verification \cite{Hoyle}. A new interaction of Nature a thousand times weaker 
than gravity with a range of about $\lambda \simeq 200~A.U. = 3.5 \times 10^{13}~m$ 
has been suggested as a solution to the inferred Pioneer 10/11 anomalous acceleration 
\cite{Anderson1}.

\item Local Position Invariance is rather poorly established in comparison to
the Weak Equivalence Principle and Local Lorentz Invariance. Clock experiments 
at the International Space Station are recommended as they represent 
the most promising way to further improve the level of 
accuracy of this symmetry \cite{Russell}.

\item Schemes to alter vacuum properties and the relative strength of the 
known fundamental interactions of Nature are beyond current theoretical 
knowledge and out of reach of forseeable technological developments 
\cite{Bertolami1}.

\end{itemize}

As far as propulsion is concerned, we have shown that if via an hypothetical scheme
gravity could be modified then, fairly independently from the underlying  
{\it modus operandi}, the following conclusions can be reached \cite{Bertolami1}:
 
\begin{itemize}

\item If modifications of local gravity were through any scheme that would  
decrease the inertial mass of a spacecraft then thrust would be lost as the 
propellant mass would decrease as well.

\item If modifications of gravity were achieved through a scheme that would affect 
the gravitational mass its impact will depend on the relative gravity contribution in 
comparison with the other components of the so-called 
$\Delta v$ requirement. For instance, to reach a Low Earth Orbit the gain 
would be rather modest. For a Geosynchronous Orbit however, the gain would be more important, but really 
significant only through a drastic ``shielding'' of the gravitational mass ($> 95 \%$).

\item Wire-like gravitomagnetic field 
assisted propulsion schemes lead to irrelevant thrust as compared to 
conventional devices such as deployable booms or electrodynamical tethers.

\item Control of gravity would bring a somewhat modest gain in terms of 
launching of spacecraft and no breakthrough for propulsion in general.

\end{itemize}

In the light of these conclusions we have suggested a set of recommendations 
to reflect our view that emphasis in what concerns the subject of gravity 
control should be focussed on microgravity, manned space flight and fundamental physics. 
In concrete terms we have strongly recommended that ESA's fundamental physics programme is
extended so to emcompass:

\noindent
{\bf 1) } Searches for violations of the Weak Equivalence Principle  
for antiparticles and charged particles at the International Space Station (ISS).
Furthermore, that the ISS is used as a platform to vigorously pursue clock comparison 
experiments.

\noindent
{\bf 2) } Studies of possible new phenomena involving the gravitomagnetic field in 
quantum materials.

\noindent
{\bf 3) } Analyses of tracking data  
of existing and forthcoming missions as well as implementing means for endowing 
future generation of spacecraft with instrumentation to reliably reconstruct 
their energy production and dissipation history.

Concerning the last point we have pointed out that a most promising strategy 
to confirm the existence of anomalous acceleration experienced by the 
Pioneer 10/11 spacecraft would involve 
a dedicated mission - provisionally named Sputnik 5 mission \cite{Bertolami1}. 
Interestingly, this view has also been recently advocated by 
the team that has firstly identified this anomaly \cite{Anderson2}.

\vspace{0.5cm}

{\large\bf Acknowledgements} 

\vspace{0.5cm}

\noindent
The authors would like to express their gratitude to  
Clovis Jacinto de Matos, ESA's technical 
officer of this study, and Jean Christophe Grenouilleau 
for their continuous support and invaluable discussions 
throught out the study. One of us (O.B.) would like also to thank 
Concei\c c\~ao Bento, Tom Girard, Filomena Nunes and 
Jorge P\'aramos for discussions and insights on various subjects 
discussed in the study.

\noindent
This study was funded by the European Space Agency General Studies Programme 
under the ESTEC Contract 15464/01/NL/SFe.

\end{document}